\documentclass[journal]{IEEEtran}

\usepackage{epsfig,amsmath,amssymb,epsf,amsthm,scalefnt, subfig}
\usepackage[table,dvipsnames]{xcolor}
\definecolor{myblue}{RGB}{0,51,160}
\definecolor{mypurple}{RGB}{142, 15, 237}
\definecolor{mygreen}{RGB}{75,148,112}
\usepackage{makecell}
\usepackage{float}
\usepackage{cite}
\usepackage{psfrag}
\usepackage{tabularx}   
\usepackage{array}      
\usepackage{siunitx}    
\usepackage{booktabs}
\usepackage{threeparttable}
\usepackage{arydshln}  
\usepackage{multirow}
\usepackage{makecell}
\usepackage{adjustbox}   
\usepackage{pifont}      
\usepackage[utf8]{inputenc}
\usepackage{enumitem}

\newcolumntype{L}[1]{>{\raggedright\arraybackslash}p{#1}}
\newcolumntype{C}[1]{>{\centering\arraybackslash}p{#1}}
\newcolumntype{Y}{>{\raggedright\arraybackslash}X}

\newtheorem*{lemma*}{Lemma}

\def\b0{{\pmb{0}}} 

\theoremstyle{remark}

\begin{document}

\title{System-Level {Evaluation} of LEO Satellite Communications Under Service-Driven Traffic Dynamics}

\author{\IEEEauthorblockN{Miyeon Lee, \textit{Graduate Student Member, IEEE}, Dong-Hyun Jung, \textit{Member, IEEE}, 
Sucheol Kim, \textit{Member, IEEE}, \\Hwanjin Kim, \textit{Member, IEEE}, and Junil Choi, \textit{Senior Member, IEEE}}\\
\thanks{Miyeon Lee and Dong-Hyun Jung contributed equally to this work.}\
\thanks{This work was supported in part by Korea Research Institute for defense Technology planning and advancement(KRIT) grant funded by the Korea government(DAPA(Defense Acquisition Program Administration))(KRIT-CT-22-040, Heterogeneous Satellite constellation based ISR Research Center, 2022), in part by the InnoCORE program of the Ministry of Science and ICT(N10250155), and in part by Basic Science Research Program through the
National Research Foundation of Korea(NRF) funded by the Ministry of Education(RS-2024-00395824).}
\thanks{Miyeon Lee and Junil Choi are with the School of Electrical Engineering, Korea Advanced Institute of Science and Technology, Daejeon, 34141, South Korea (e-mail: \{mylee0031; junil\}@kaist.ac.kr).}
\thanks{Dong-Hyun Jung is with the School of Electronic Engineering, Soongsil University, Seoul, 06978, South Korea (e-mail: dhjung@ssu.ac.kr).}
\thanks{Sucheol Kim is with the Satellite Communication Research
Division, Electronics and Telecommunications Research Institute, Daejeon, 34129, South Korea (e-mail: loehcusmik@etri.re.kr).}
\thanks{Hwanjin Kim is with the School of Electronics Engineering, Kyungpook National
University, Daegu 41566, South Korea (e-mail: hwanjin@knu.ac.kr).}}

\maketitle
\begin{abstract} As low Earth orbit (LEO) satellite communications take shape on a global scale, system-level evaluation has become essential for a rigorous understanding of their performance characteristics. Most existing studies have relied on a full-buffer assumption, which obscures the heterogeneity and intermittency that characterize actual user equipment (UE) traffic. This article evaluates the performance of LEO satellite communications under realistic service conditions through the incorporation of non-full-buffer traffic models. To this end, we develop a system-level simulator that complies with the channel modeling and evaluation methodologies specified in 3rd Generation Partnership Project (3GPP) technical reports 38.811 and 38.821. Three key system-level performance metrics are considered: UE throughput, resource block (RB) allocation ratio, and packet delay, which are evaluated across 3GPP LEO satellite study cases under diverse traffic models. The throughput results characterize the distribution of achievable data-rates, which delineates the practical operating boundaries of the system. The RB allocation analysis reveals patterns of resource consumption as UE density varies, which provides a quantitative basis for assessing system capability. Furthermore, the delay analysis characterizes latency behavior, which is of particular importance in satellite environments where substantial propagation delays are inherent and must be examined to ensure service feasibility.
These evaluations provide a realistic performance outlook for non-full-buffer LEO satellite communications and provide insight into the user experience under practical operating conditions.
\end{abstract}

\section{Introduction}\label{sec1}
Satellite communications are poised to become a cornerstone of 6G wireless systems, serving as a key enabler of ubiquitous connectivity. Despite the rapid global expansion of 5G networks, a significant digital divide remains. Low Earth orbit (LEO) satellite constellations have been deployed to address this disparity by offering global coverage at significantly lower infrastructure costs than terrestrial-only network expansion.

Interest in LEO satellite systems is driven by both commercial viability and operational advantages. Commercially, the industry has shifted toward cost-efficient manufacturing and shortened lead times, with compact satellite platforms significantly reducing launch expenditures. Operationally, LEO satellites operate at altitudes between 300~km and 2,000~km, enabling lower propagation loss and reduced latency compared to geostationary orbit (GEO) satellites at 35,786~km. While GEO satellite communications offer broader coverage, their inherent latency limits support for delay-sensitive services. Medium Earth orbit (MEO) satellites occupy an intermediate zone but remain largely limited to maritime platforms, aeronautical communication, and specialized backhaul applications due to stringent antenna and terminal power requirements. Consequently, investment momentum has increasingly concentrated on LEO satellite communications, including private deployments such as SpaceX's Starlink and OneWeb, as well as government-supported initiatives such as the European Union's infrastructure for resilience, interconnection and security by satellites ($\text{IRIS}^2$) program and China's GuoWang constellation \cite{nature}. These efforts reflect a growing consensus that LEO satellite communications offer a favorable balance among coverage, latency, and deployability.

\begin{figure*} [t]
	\centering
	\includegraphics[width=\textwidth, height=0.38\textheight, keepaspectratio]{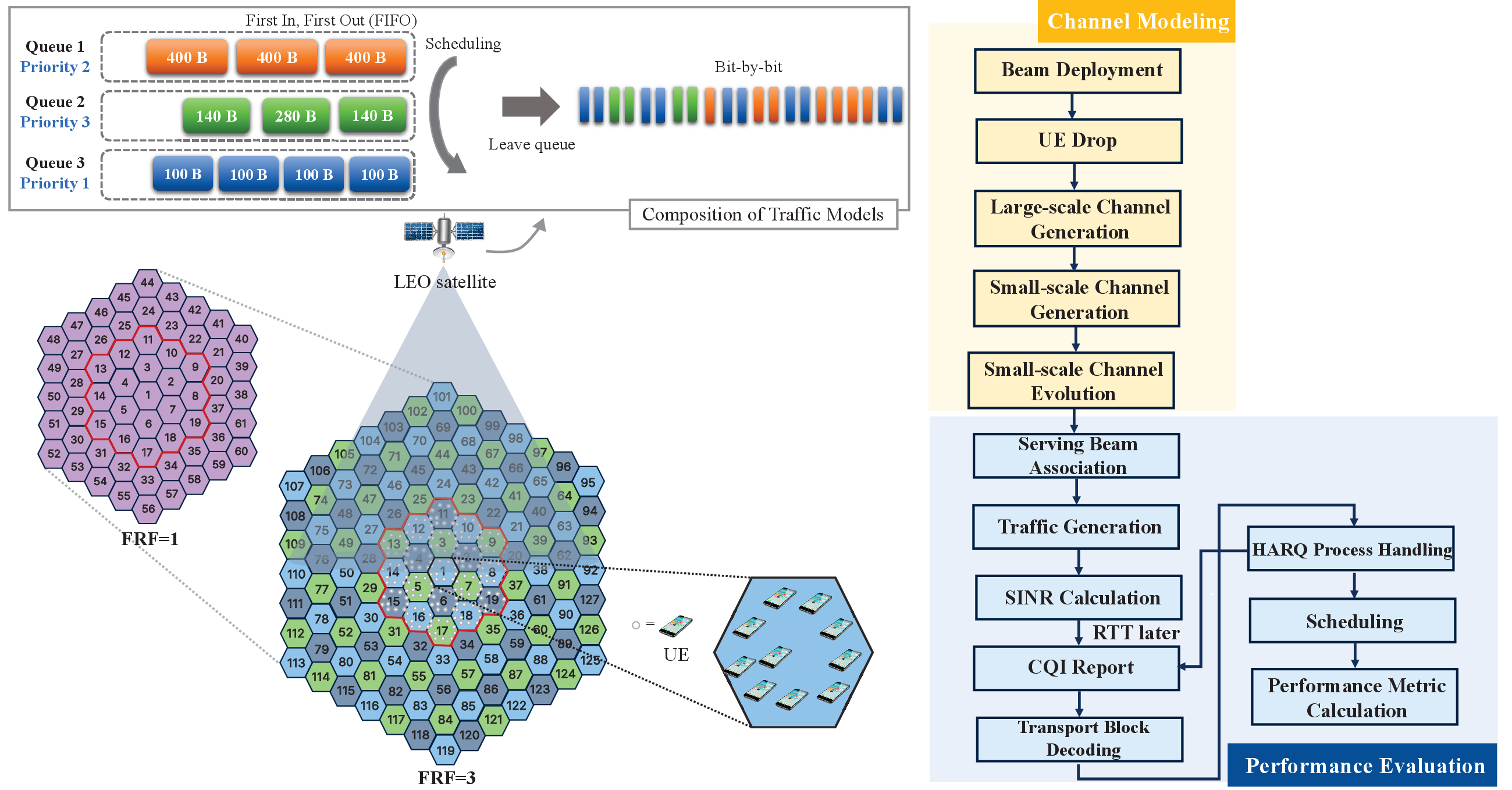}
	\caption{System-level simulation methodology, presenting the system model (left) and the procedural flow (right).}\label{system}
\end{figure*}

Despite these advantages, the technical challenges inherent to LEO satellite communication systems necessitate a rigorous and comprehensive performance assessment. Such an assessment extends beyond verification of link budget sufficiency, which ensures adequate received power at the antenna front end \cite{Guidotti}, and must also establish whether performance objectives can be met under multi-user equipment (UE) operation across wide service areas. To enable this evaluation, a hierarchical simulation framework that integrates link-level and system-level simulations is employed \cite{klaus}. Link-level simulation evaluates the end-to-end physical-layer processing chain, including modulation, channel coding, and receiver algorithms, to quantify the achievable performance of a single communication link. However, this does not capture the interactions between multiple UEs that contend for shared time and frequency resources. System-level simulation addresses this limitation by modeling resource management and traffic dynamics across multiple beams and UEs, thereby enabling evaluation of network-wide metrics such as throughput and quality of service (QoS) \cite{Lichen, Jao}. Within this hierarchical framework, link-level simulation is used to derive performance models that map physical-layer operating conditions to link performance metrics. These models are then incorporated into the system-level simulation to reflect physical-layer effects without explicit waveform-level processing, which maintains physical-layer fidelity while enabling scalable evaluation of multi-user interactions and higher-layer protocols.


A standardized framework for system-level evaluation of LEO satellite communications has been established by the 3rd Generation Partnership Project (3GPP) through Release~15 technical report (TR)~38.811 \cite{TR38811} and its extension in Release~16 TR~38.821 \cite{TR38821}. TR~38.811 specifies satellite geometry, mobility, Doppler characteristics, propagation delay models, and key channel components, while TR~38.821 defines calibration assumptions and beam layouts and provides calibration study cases that span variations in orbital altitude, satellite parameters, beam elevation, frequency band, and frequency reuse factor (FRF). Together, these documents form the primary reference framework for reproducible and methodologically consistent system-level simulation.


In operational deployments, LEO satellite systems must be evaluated under both throughput and latency requirements for delay-sensitive services. However, most system-level simulation studies of LEO satellite communications reported to date have relied on a full-buffer traffic model, including analyses of beam hopping \cite{zhang} and performance evaluation based on ITU requirements \cite{magister}. Although this assumption simplifies analysis, it fails to represent the bursty and intermittent nature of practical traffic and therefore obscures queueing behavior and scheduling delays. {Existing packet-level simulation frameworks, including SNS3~\cite{sns3} and Hypatia~\cite{Hypatia}, provide capabilities for non-full-buffer traffic analysis. However, SNS3 is mainly oriented toward Digital Video Broadcasting Satellite Second Generation (DVB-S2)/2nd Generation Interactive DVB Satellite System (DVB-RCS2)-based satellite scenarios and long range (LoRa)-based Internet of Things (IoT) scenarios, whereas Hypatia focuses on Transmission Control Protocol (TCP)/User Datagram Protocol (UDP) flows over dynamic LEO network topologies.}

{This article addresses this lacuna by developing a 3GPP-compliant system-level evaluation framework for LEO satellite communications under service-driven non-full-buffer traffic models, incorporating stochastic traffic processes that capture packet inter-arrival time variability, packet size diversity, and intermittent activity patterns.} The study focuses on the {prioritized calibration study cases~9 and~10} defined in TR~38.821, which correspond to S-band handheld terminals with FRF~$=1$ and FRF~$=3$, respectively. System performance is assessed in terms of UE throughput, resource block (RB) allocation ratio, and packet delay.


\section{Overview of System-Level Simulator}\label{sec2}
Following the design principles specified in TR~38.811 and~38.821, the system-level simulator for LEO satellite communications is organized into a set of interdependent modules that include the beam layout model, the channel model, the stochastic traffic model, and auxiliary components for performance evaluation. This section explains the modeling assumptions that underlie these modules and describes their implementation within our proprietary MATLAB-based system-level simulator.

\subsection{Beam Layout Model}
The methodologies specified in TR~38.821 describe beam layout structures that are determined by the FRF, with FRF~$=1$ and FRF~$=3$, as illustrated in Fig.~\ref{system}. These layouts provide the basis for a range of service beam configurations that commonly include single-beam, 7-beam, 19-beam, and 36-beam topologies. An increase in the number of service beams improves the fidelity of inter-beam interference (IBI) representation, as it could generate realistic interference from neighboring beams. However, too large number of beams entails a substantial increase in simulation complexity, which arises from the repeated configuration and execution that system-level calibration requires. By contrast, configurations that rely on a small number of service beams reduce computational burden but fail to capture the interference experienced by UEs, particularly in outer beams, which have fewer neighboring beams than those in the inner region. To address this limitation, wraparound techniques that follow the methodology described in TR~38.821 are employed. Through this approach, beams within the central cluster experience interference comparable to those in large-scale deployments, without requiring an excessively large simulation domain.

\subsection{Channel Model}
The fast fading model specified in TR~38.901 \cite{TR38901} can be extended to LEO satellite communication scenarios through appropriate adjustments of cluster parameters. Compared to terrestrial environments, satellite channels exhibit substantially smaller multipath clusters due to line-of-sight propagation dominance and sparse local scatterers. To account for these characteristics, angular spread scaling is applied during cluster generation, through which the resulting clusters represent the reduced scattering conditions that are intrinsic to LEO satellite channels.

TR~38.811 further specifies channel parameters for LEO satellite communications, which include delay spread, angle spread, shadow fading, and the Rician $K$-factors for dense urban, urban, suburban, and rural environments. In addition to these parameters, several propagation effects are incorporated to obtain an accurate channel representation. Phase rotation that arises from time-varying Doppler shift is applied to each ray, as both the satellite and the UE exhibit time-dependent motion. Faraday rotation is also incorporated to capture the rotation of the polarization plane that arises when the electromagnetic wave propagates through the ionized medium in the presence of Earth's magnetic field. This effect is applied to the channel coefficients of all rays within each cluster, which reflects the polarization changes introduced by the ionosphere along spaceborne communication paths. 

\subsection{Non-full-buffer Traffic Models}
As LEO satellite communication services expand and their application domains diversify, traffic models that capture heterogeneous service characteristics have become essential. To reflect such diversity within a system-level simulation framework, four stochastic traffic models are defined, each corresponding to a distinct service pattern.

The first two traffic models are based on Poisson processes with fixed and random packet sizes, respectively. {A homogeneous Poisson process is adopted to model sporadic short-packet services, such as small web-object downloads, IoT control messages, and text-based short messages.} Although highly bursty services may require non-homogeneous Poisson processes, the homogeneous Poisson process captures random packet arrivals and the memoryless nature of these services through exponentially distributed inter-arrival times. This traffic modeling choice is consistent with 3GPP evaluation methodologies, including the file transfer protocol model specified in TR 36.814 \cite{TR36814}. Within this framework, packet sizes are treated as either fixed or variable, depending on the service characteristics. Fixed-size packets represent small web-object downloads and IoT control messages, whereas variable-size packets provide a more appropriate abstraction for {text-based short messages}. This model enables assessment of throughput stability, scheduling efficiency, and packet delay under sporadic packet arrivals across different packet size regimes.

The third traffic model is defined as a two-state Markov-modulated on/off process that represents delay-sensitive services such as group call scenarios. This stochastic formulation supports an evaluation of packet delay behavior, as well as throughput stability and scheduling efficiency, under conversational traffic conditions. In a group call scenario, multiple users alternate between talkspurt (on) and listening or silence (off) states, where the on and off durations can be modeled by exponential distributions. During a talkspurt, the active talker generates uplink (UL) voice packets, while users in the off state receive constant-bit-rate downlink (DL) packets that are generated every 20 ms by standard voice codecs. Since only one user can speak at a time, an attempt to transition from the off state to the on state is blocked when another user is speaking and is rescheduled after a newly sampled silence interval. This structure reflects the contention dynamics and talkspurt behavior observed in conversational voice traffic.

The final traffic model is a mixed traffic configuration that combines Poisson processes with fixed and variable packet sizes and a Markov-modulated process, which reflects practical network environments in which a single system simultaneously supports multiple services with distinct traffic characteristics. In this model, the three previously described traffic generators operate concurrently, and packets that originate from different services are enqueued in per-service queues that follow a first-in-first-out manner, while scheduling across queues follows a strict priority policy, as illustrated in Fig.~\ref{system}. Under strict priority scheduling, packets in lower-priority queues are transmitted only after all packets in higher-priority queues have been served. Consequently, lower-priority queues may experience starvation, i.e., they receive no service, when higher-priority queues remain persistently backlogged. To mitigate this effect, a starvation threshold is introduced to ensure that packet delays in low-priority queues remain within acceptable limits.

In contrast to the full-buffer assumption, which presumes infinite backlog and ignores packet arrival timing, the considered traffic models introduce application-driven load variations that reflect practical service behavior. By incorporating packet arrival patterns and state-dependent activity processes, the simulator enables an evaluation of QoS metrics, including packet delay, that cannot be captured under the full-buffer traffic assumption.

\subsection{Evaluation Framework}
To quantify the system-level performance, a time-driven simulation framework that is consistent with 3GPP methodology is employed, as illustrated on the right side of Fig.~\ref{system}. The simulator is designed to represent the time evolution of the DL channel, traffic arrivals, and scheduling decisions, so that the combined effects of physical-layer behavior and higher-layer mechanisms can be examined within a unified framework.

At each simulation step, the instantaneous link quality at the RB level is determined by time-varying channel conditions, which include large-scale path loss, small-scale fading, and IBI. Instead of performing full physical-layer decoding, standard physical-layer abstraction techniques are applied, through which RB-level channel conditions are mapped to an effective measure of link reliability. Channel quality indicator (CQI) feedback that is subject to round-trip time (RTT) delay is then used to guide transmission decisions at the satellite, which reflects the delayed feedback inherent in LEO satellite systems.

In parallel, stochastic non-full-buffer traffic is modeled to represent downlink service demand, resulting in time-varying queue occupancy at the LEO satellite scheduler for each UE and introducing queueing and delay dynamics. RBs are allocated through a proportional fair (PF) scheduling process that balances instantaneous channel opportunities with long-term fairness across UEs. In addition, hybrid automatic repeat request (HARQ) mechanisms are employed to ensure robustness against decoding errors through retransmissions. These components jointly form a closed-loop process in which channel variations, traffic dynamics, scheduling decisions, and retransmission events continuously influence one another over time.


System-level performance metrics are derived from these cross-layer interactions. Throughput is computed by accumulating decoded transport blocks over the simulation time. The simulator also records the RB allocation ratio, which is defined as the fraction of allocated RBs relative to the total available RBs and serves as an indicator of spectrum utilization. Packet delay is measured as the elapsed time between packet arrival at the scheduler-side queue for each UE at the LEO satellite and successful decoding at the UE, which captures the combined effects of queueing, scheduling, HARQ retransmissions, and RTT.


\begin{table}[t]
\centering
\renewcommand{\arraystretch}{1.25}
\setlength{\arrayrulewidth}{0.7pt}  
\caption{{Simulation configurations.}}\label{table_parameters}
\begin{tabularx}{\linewidth}{
  |>{\raggedright\arraybackslash}X|
   >{\raggedright\arraybackslash}X|}
\hline
\rowcolor{mygreen!60}
\textbf{Parameters} & \textbf{Value} \\
\hline

\rowcolor{mygreen!30}
\multicolumn{2}{|l|}{\textbf{Satellite parameter}} \\
\hline
Altitude                           & \SI{600}{km} (LEO)\\
Payload                            & Transparent payload\\
Frequency band                     & S-band (\SI{2}{GHz}) \\
Channel bandwidth                  & \SI{30}{MHz} \\
{Subcarrier spacing}               & \SI{15} {kHz}\\
{Maximum antenna gain}         & \SI{30}{dBi} \\
3 dB beam width                    & 4.4127$^\circ$ \\
EIRP density                       & \SI{34}{dBW/MHz} \\
G/T                                & \SI{1.1}{dB/K} \\
Antenna aperture                   & \SI{2}{m}\\
Antenna polarization configuration & Circular \\
{Central beam elevation}    & $90^\circ$ \\
{Antenna pattern}                    & {Sec. 6.4.1 in TR~38.811} \\
\hline

\rowcolor{mygreen!30}
\multicolumn{2}{|l|}{\textbf{UE parameter}} \\
\hline
Configuration          & Handheld \\
Antenna polarization   & Linear: $\pm45^\circ$-pol \\
Antenna gain           & \SI{0}{dBi} \\
Noise figure           & \SI{7}{dB} \\
Antenna configuration  & (1, 1, 2) with omni-directional antenna\\
Distribution           & Outdoor; 10 UEs/beam\\
Environment            & Rural\\
Velocity               & \SI{3}{km/h} \\
\hline

\rowcolor{mygreen!30}
\multicolumn{2}{|l|}{\textbf{Channel parameter}} \\
\hline
Propagation conditions & Clear sky, line-of-sight\\
Large-scale fading          & Sec. 6.6 in TR 38.811\\
Small-scale fading          & Table 6.1.1.1-7 in TR 38.821\\
Scintillation loss     & \SI{2.2}{dB} \\
\hline

\rowcolor{mygreen!30}
\multicolumn{2}{|l|}{\textbf{Simulation settings}} \\
\hline
FRF & 1 (study case~9) or 3 (study case~10) \\
Wraparound & {Two-tiers} (FRF~$=1$) or four-tiers (FRF~$=3$)\\
UE attachment           & {Based on reference} signal received power \\
Receiver type           & Minimum mean square error{-}interference rejection combining \\ 
{Processing and scheduling delay} & 3~ms\\
HARQ                    & Enabled \\
Scheduler               & {PF} \\
Number of UE drops        & 5 \\
\hline
\end{tabularx}
\end{table}

\newcolumntype{M}[1]{>{\centering\arraybackslash}m{#1}}

\begin{table*}[t]
\centering
\renewcommand{\arraystretch}{1.3}
\caption{{Traffic models with scenario-specific packet size and generation characteristics.}}\label{table_traffic}
\begin{adjustbox}{max width=\textwidth}
\begin{tabular}{M{1.7cm}|M{3.9cm}|M{2.3cm}|M{2.3cm}|M{1.1cm}|M{4.0cm}}
\specialrule{1.1pt}{0pt}{0pt}

\rowcolor{mygreen!60}

\textbf{Traffic model}
& \textbf{Traffic type}
& \multicolumn{2}{c|}{\textbf{Packet size ({Bytes})}}
& \textbf{IAT (ms)}
& \textbf{Generation logic} \\[-1pt]

\cline{3-4}
\rowcolor{mygreen!60}
& 
& \textbf{Scenario 1} 
& \textbf{Scenario 2} 
&
&\\
\specialrule{1.1pt}{0pt}{0pt}
1 & Poisson process & 400 & 800 & 100 & Fixed packet size\\
\hline

\multirow{2}{*}{2} &
\multirow{2}{*}{Poisson process} &
\multicolumn{2}{c|}{140 per segment with $K_{\mathrm{max}}=3$} &
\multirow{2}{*}{200} &
\multirow{2}{*}{Variable packet size} \\
\cline{3-4}
& & {$P(K)=\{0.8,\,0.1,\,0.1\}$} & {$P(K)=\{0.1,\,0.8,\,0.1\}$} & & \\
\hline

3 & Markov-modulated on/off process & 100 & 200 & 20 & $T_{\mathrm{on}}=2$~s,\ $T_{\mathrm{off}}=1$~s  \\
\hline

4 & Composite of traffic models {1-3} & \multicolumn{3}{c|}{\textit{Integration of scenario-specific parameters}} & {Priority: Traffic model~3 $>$ Traffic model~1 $>$ Traffic model~2} \\
\hline

\specialrule{0.9pt}{0pt}{0pt}

\end{tabular}
\end{adjustbox}
\begin{tablenotes} 
\footnotesize \item Note: IAT denotes inter-arrival time. For traffic model 2, $P(K)$ represents the probability distribution $\{P(K=1), P(K=2), P(K=3)\}$. 
\end{tablenotes}


\end{table*}

\section{System-Level Simulation Setups}
In this section, we describe the simulation configurations and assumptions adopted in our system-level simulator for LEO satellite communications. The setup follows the calibration study cases defined in 3GPP TR~38.821, with a specific focus on study cases~9 and~10 that differ in FRF value of 1 and 3, respectively, and are designated as high-priority scenarios for system-level performance evaluation.

The deployment scenario considered in this article corresponds to a direct access satellite-to-handheld link in the S-band at 2 GHz, a configuration widely regarded as one of the most prominent broadband use cases for 3GPP non-terrestrial networks. As depicted on the left side of Fig.~\ref{system}, the system consists of a multi-beam LEO satellite serving UEs distributed within the coverage areas of the individual beams. Each beam serves 10 UEs uniformly distributed within its footprint. The central 19 beams are designated as the region of interest for evaluating the statistical system performance. 

To address the underestimation of IBI, which is particularly pronounced for outer beams, a wraparound technique is employed to introduce interference contributions from neighboring beams. The number of surrounding beam tiers is determined by the FRF. For the FRF~$=1$ configuration, a two-tier wraparound structure is applied, in which the central 19 beams are surrounded by two layers of 42 beams, resulting in a total of 61 beams. For the FRF~$=3$ configuration, a four-tier wraparound structure is adopted, where 108 additional co-channel beams reusing the same frequency bands are arranged in four tiers around the central 19 beams, yielding a total of 127 beams. The remaining simulation parameters are aligned with the system-level calibration specifications in TR~38.821, which include satellite, UE, and channel characteristics, as well as key simulation assumptions. These parameters are summarized in Table~\ref{table_parameters}. 

To examine the system-level performance implications of non-full-buffer traffic, four stochastic traffic models are considered under two scenarios that differ in packet size, as summarized in Table~\ref{table_traffic}. Traffic model~1 is based on a Poisson process with fixed packet sizes of 400~bytes in Scenario~1 and 800~bytes in Scenario~2.
Traffic model~2 is also based on a Poisson process but incorporates variable packet sizes. The traffic is characterized by small packet sizes and infrequent arrivals, which lead to low data rate requirements. Each packet comprises a random number of 140~bytes segments ranging from 1 to a maximum value denoted by $K_\mathrm{max}$, drawn from a discrete non-uniform distribution specified for per scenario. This yields expected packet sizes of 182~bytes in Scenario~1 and 280~bytes in Scenario~2. Traffic model~3 employs a Markov-modulated on/off process to emulate conversational voice traffic, where the on and off periods correspond to talkspurt and silence durations modeled as exponentially distributed random variables with mean duration $T_{\mathrm{on}}$ and $T_{\mathrm{off}}$, respectively. Voice packets of 100~bytes in Scenario~1 and 200~bytes in Scenario~2 are generated accordingly. Traffic model~4 represents a composite traffic model that integrates the preceding three traffic models within each scenario while accounting for service-level priority awareness.

The analysis presented in this article concentrates on DL transmissions, as UE-driven satellite services typically exhibit highly asymmetric traffic patterns dominated by DL data volumes, making DL performance the primary determinant of UE-perceived QoS. This modeling choice is consistent with prior system-level simulation studies in terrestrial networks that have similarly emphasized DL performance evaluation \cite{bjorn}. From a technical standpoint, these DL findings cannot be directly extended to UL, as DL-UL reciprocity does not generally hold in LEO satellite communications due to asymmetries in link budgets, frequency division duplexing, Doppler dynamics, antenna constraints, and terminal power availability. Therefore, a detailed assessment of UL performance is deferred to future work.

\begin{figure*}[t]
  \centering
  \captionsetup[subfloat]{labelformat=empty}

  \subfloat[(a) Traffic model~1 (study case~9)\label{fig:six-a}]{
    \includegraphics[width=0.31\textwidth]{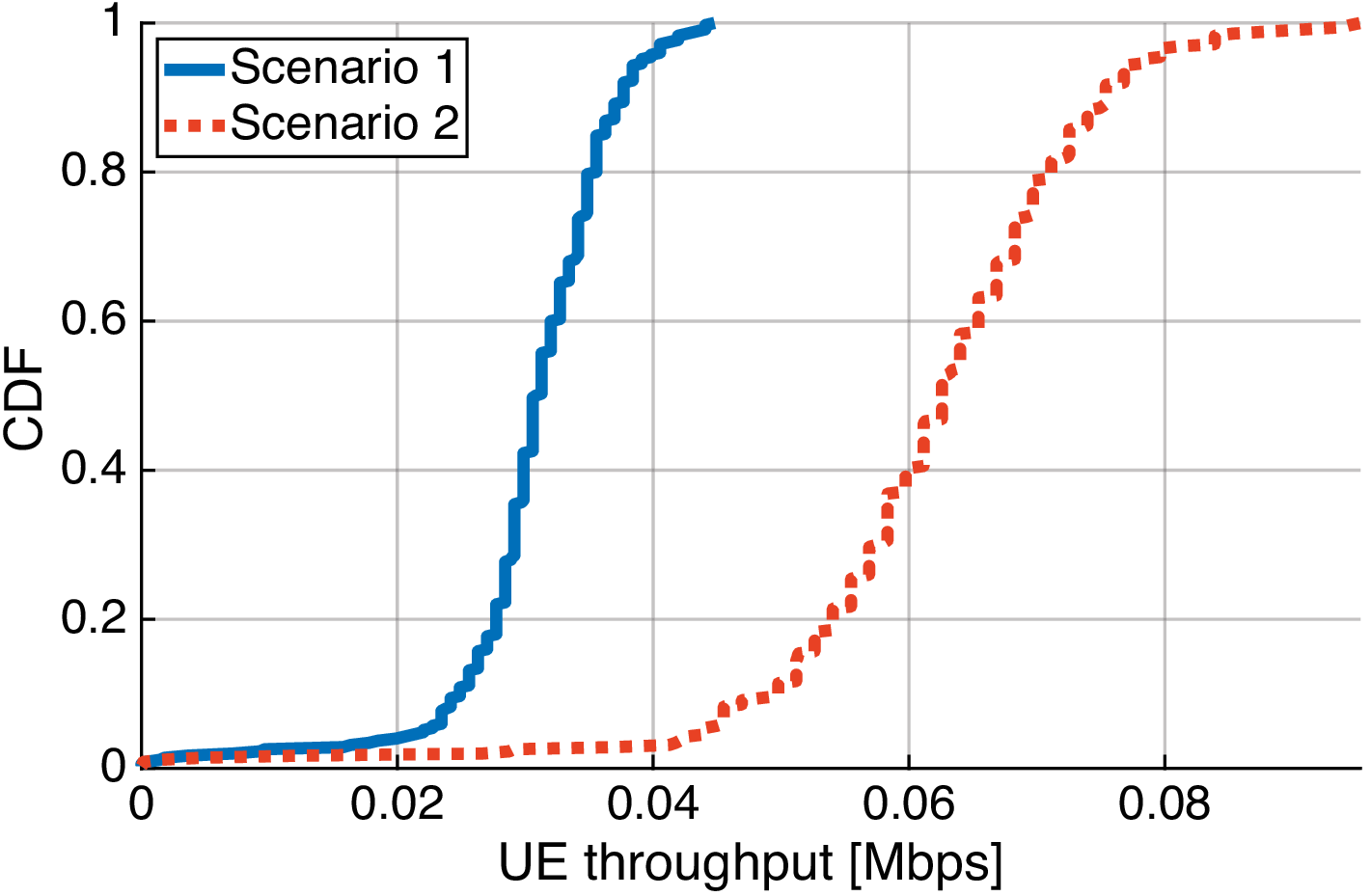}
  }\hfill
  \subfloat[(c) Traffic model~2 (study case~9)\label{fig:six-c}]{
    \includegraphics[width=0.31\textwidth]{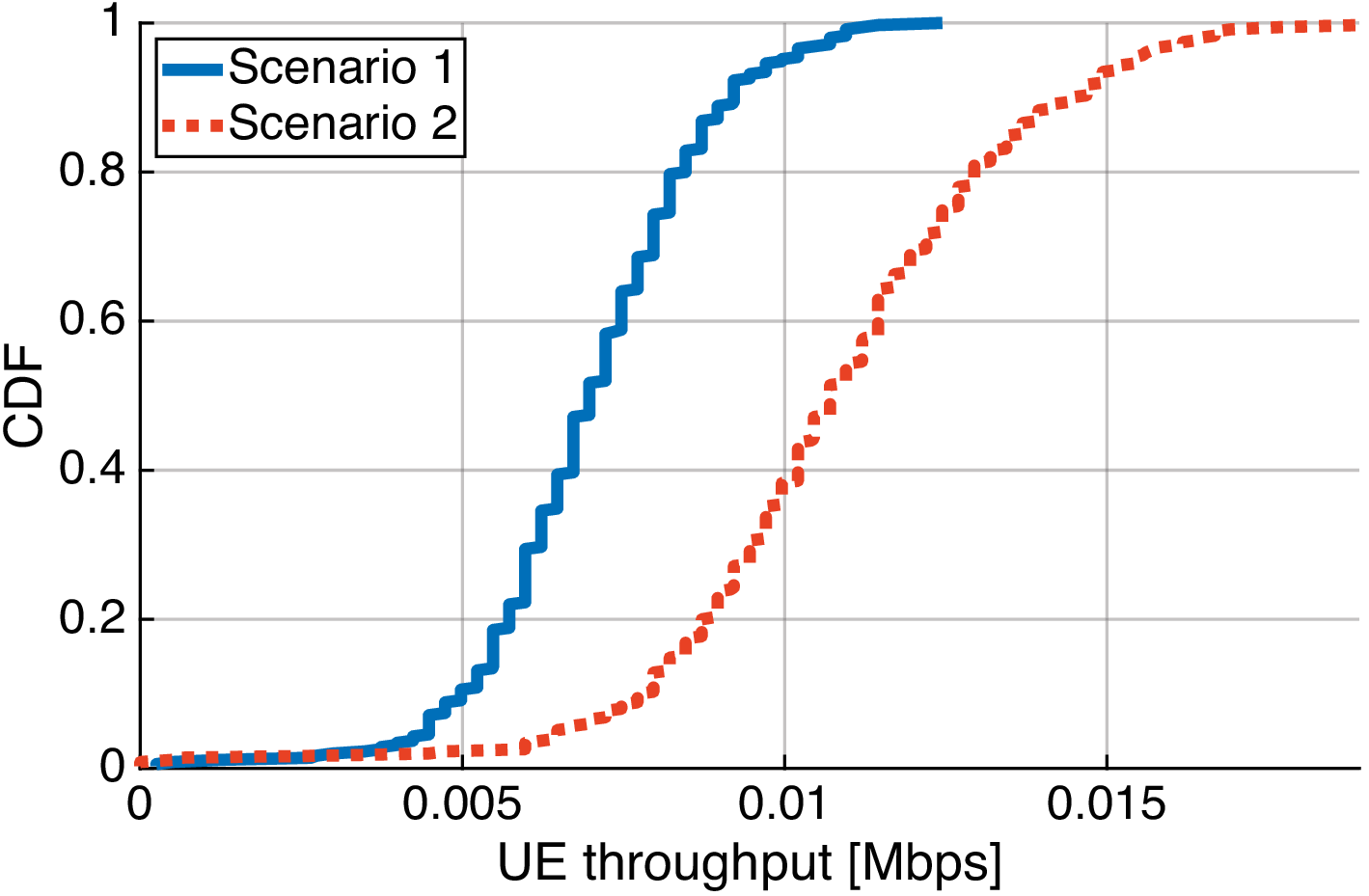}
  }\hfill
  \subfloat[(e) Traffic model~3 (study case~9)\label{fig:six-e}]{
    \includegraphics[width=0.31\textwidth]{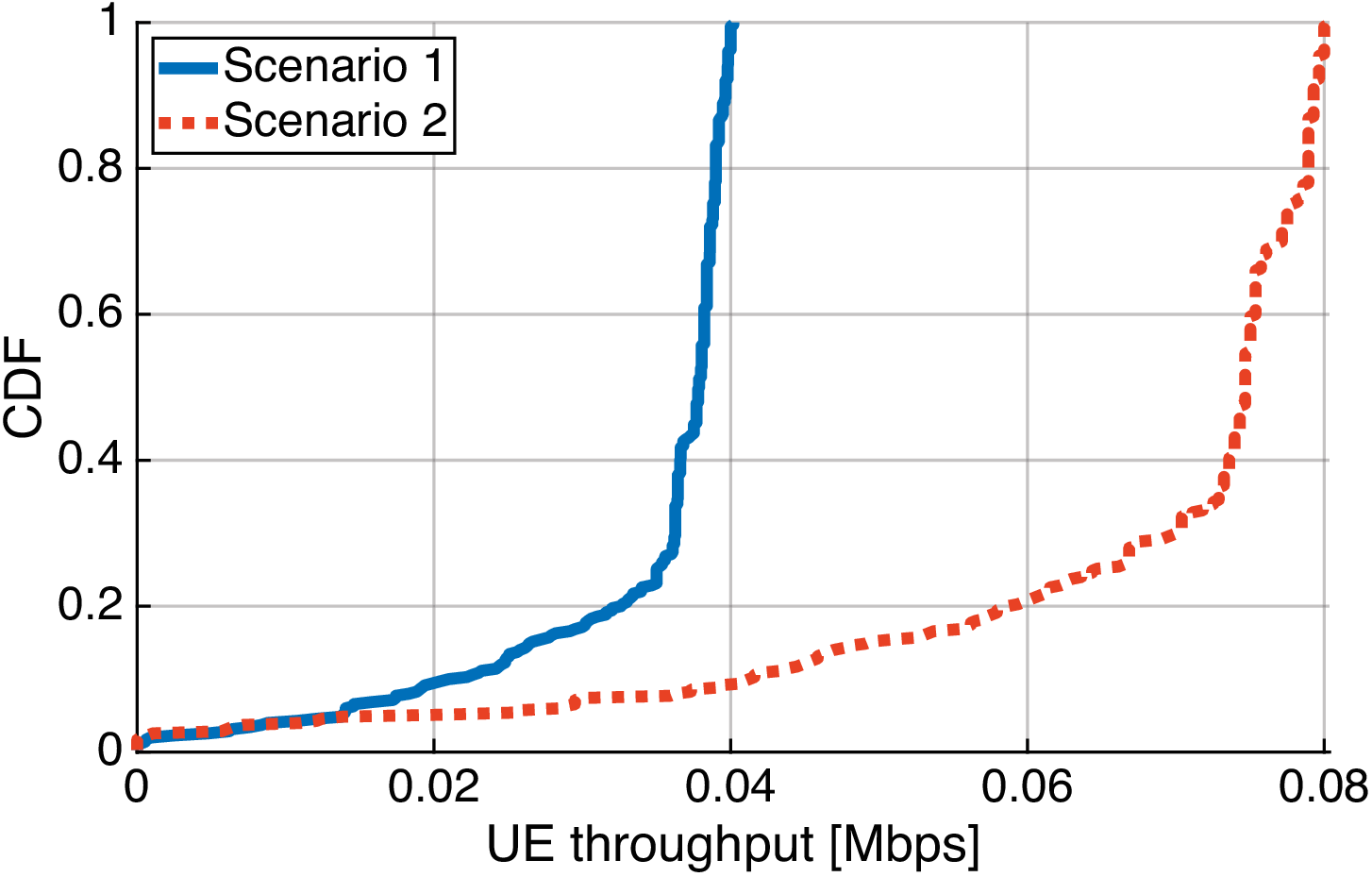}
  }

  \subfloat[(b) Traffic model~1 (study case~10)\label{fig:six-b}]{
    \includegraphics[width=0.31\textwidth]{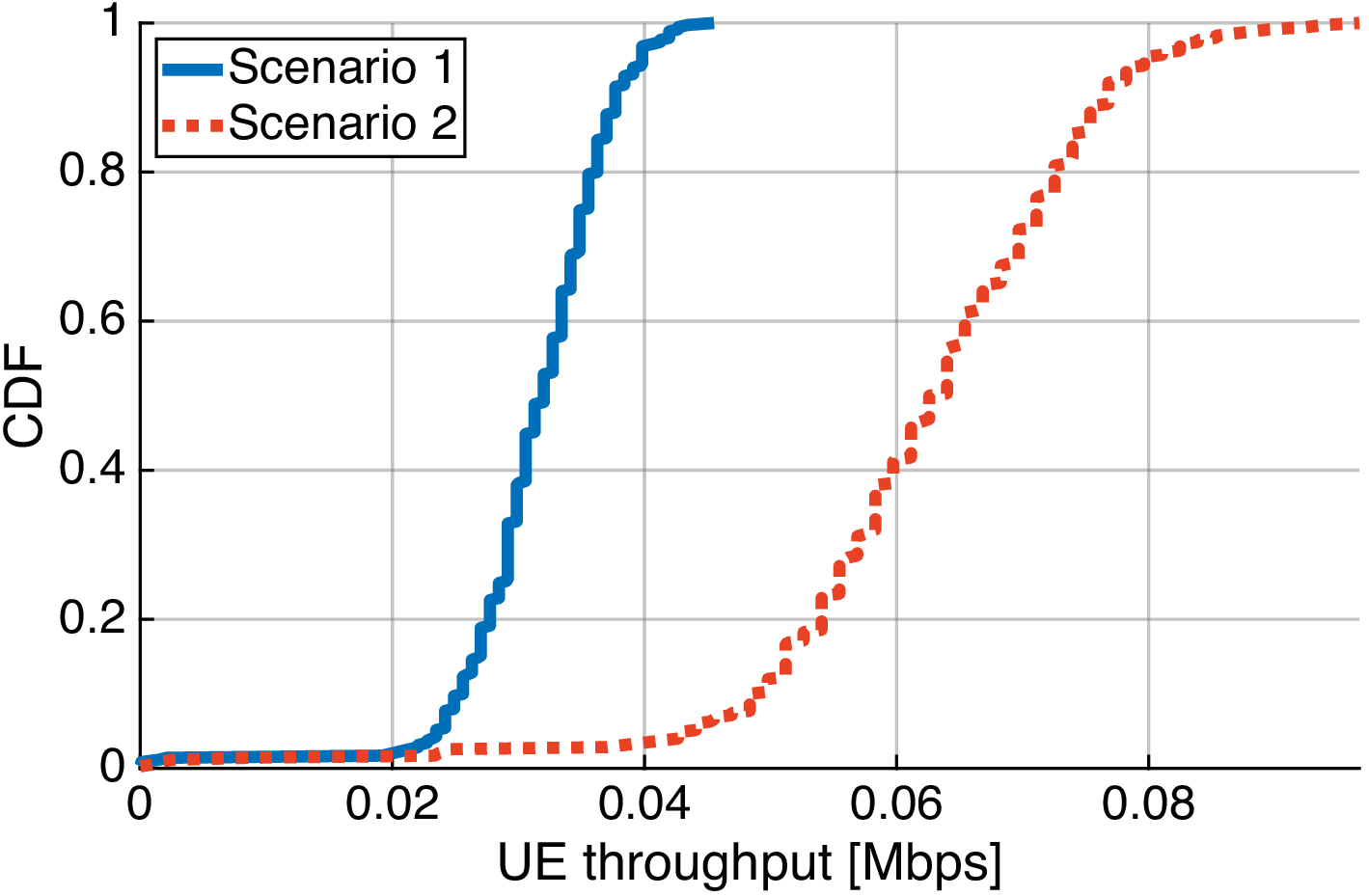}
  }\hfill
  \subfloat[(d) Traffic model~2 (study case~10)\label{fig:six-d}]{
    \includegraphics[width=0.31\textwidth]{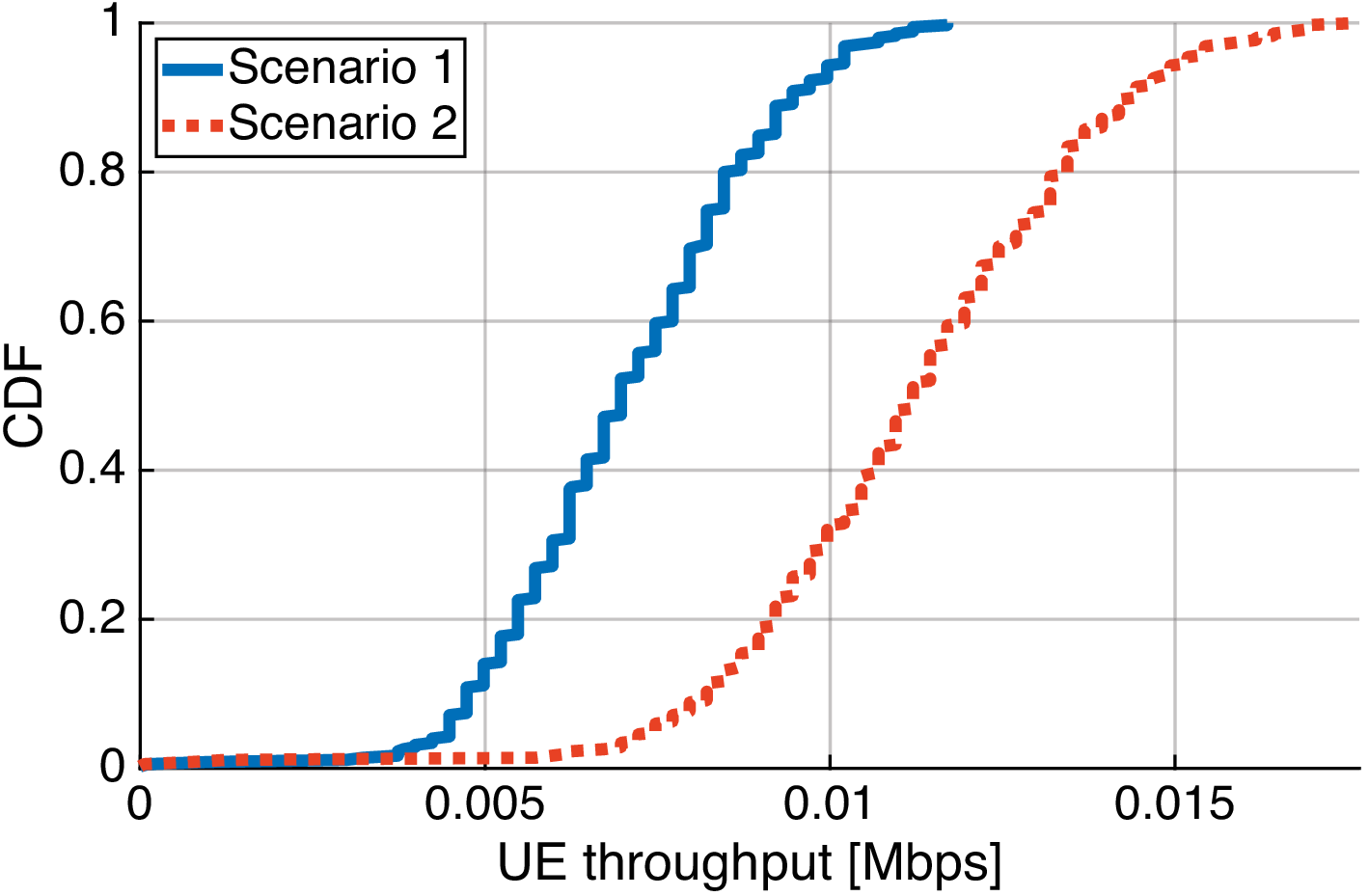}
  }\hfill
  \subfloat[(f) Traffic model~3 (study case~10)\label{fig:six-f}]{
    \includegraphics[width=0.31\textwidth]{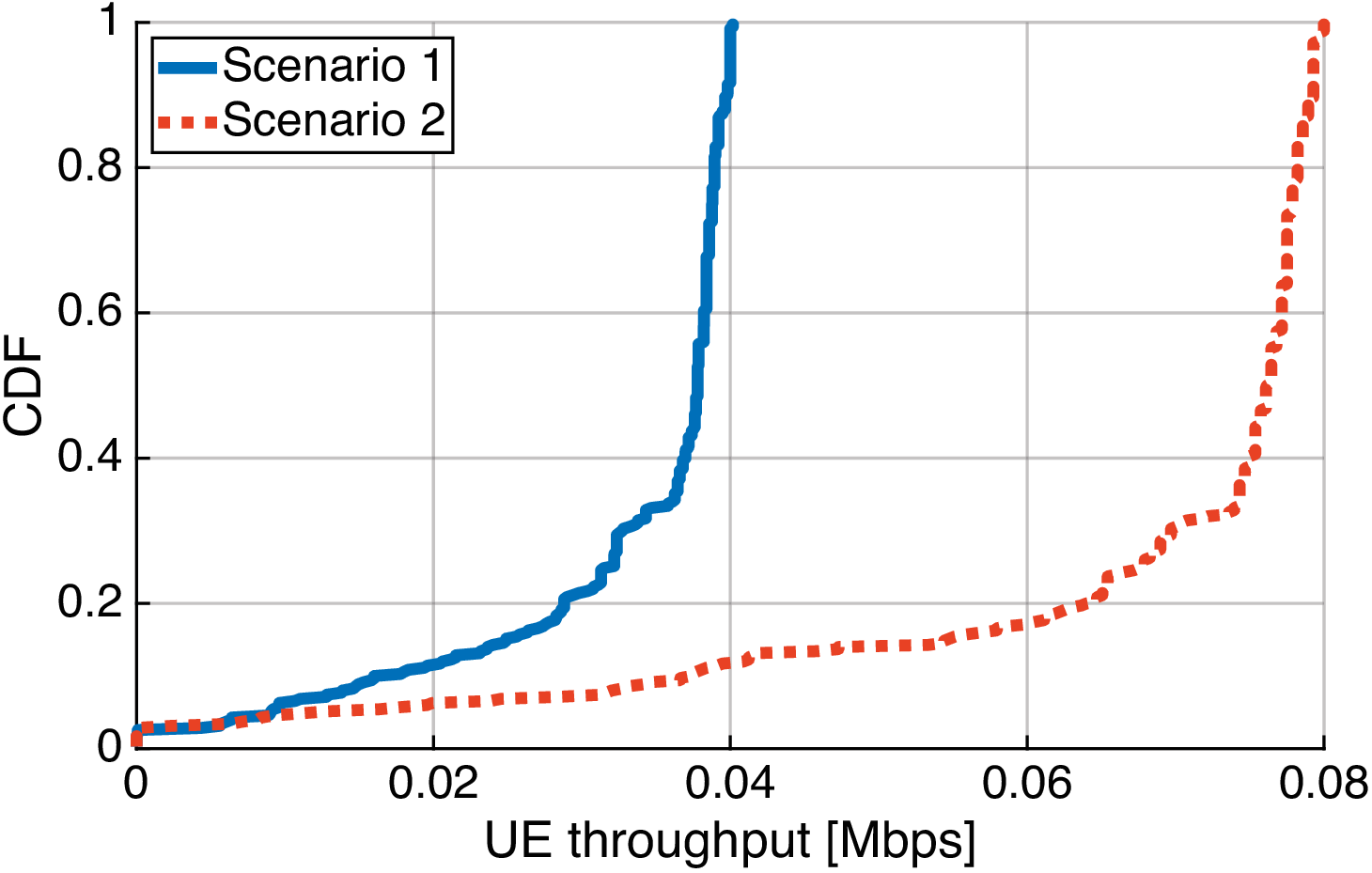}
  }
  \caption{UE throughput in study cases 9 and 10 for Scenarios 1 and 2 under traffic models 1 (left), 2 (center), and 3 (right).}
  \label{fig_tm1to3}
\end{figure*}

\section{System-Level Simulation Results}
In this section, we present the outcomes obtained using the system-level simulator. Four non-full-buffer traffic models are evaluated under study cases~9 and~10, with a focus on UE throughput, RB allocation ratio, and packet delay. Additionally, two scenarios with different packet-size configurations are considered to examine the impact of packet size on throughput for each traffic model, as summarized in Table~\ref{table_traffic}.

\begin{figure}[t]
  \centering
  \subfloat[Traffic model 4 (study case~9)\label{fig:top}]{
    \includegraphics[width=0.72\linewidth]{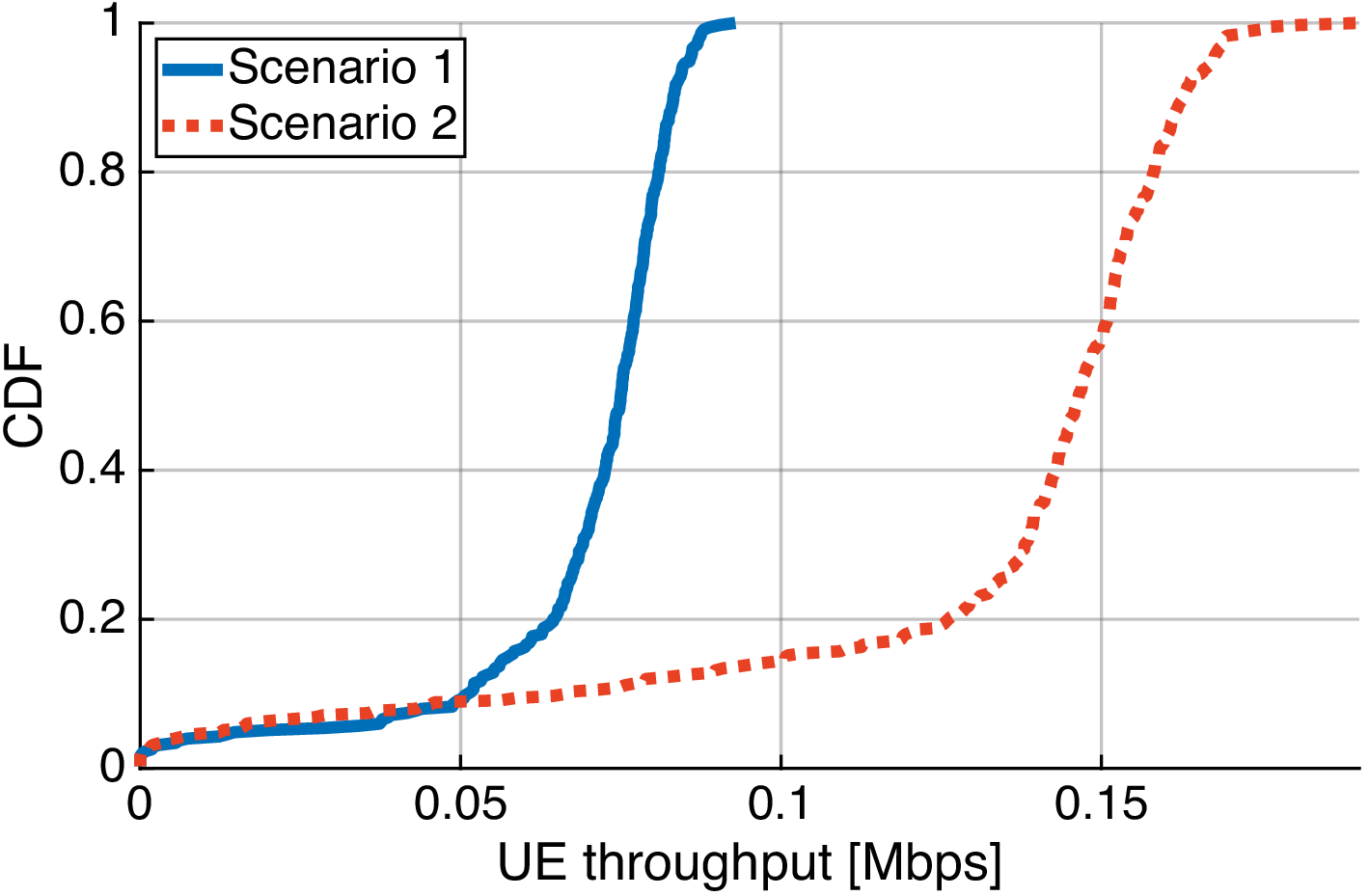}
  }

  \medskip 

  \subfloat[Traffic model 4 (study case~10)\label{fig:bottom}]{
    \includegraphics[width=0.72\linewidth]{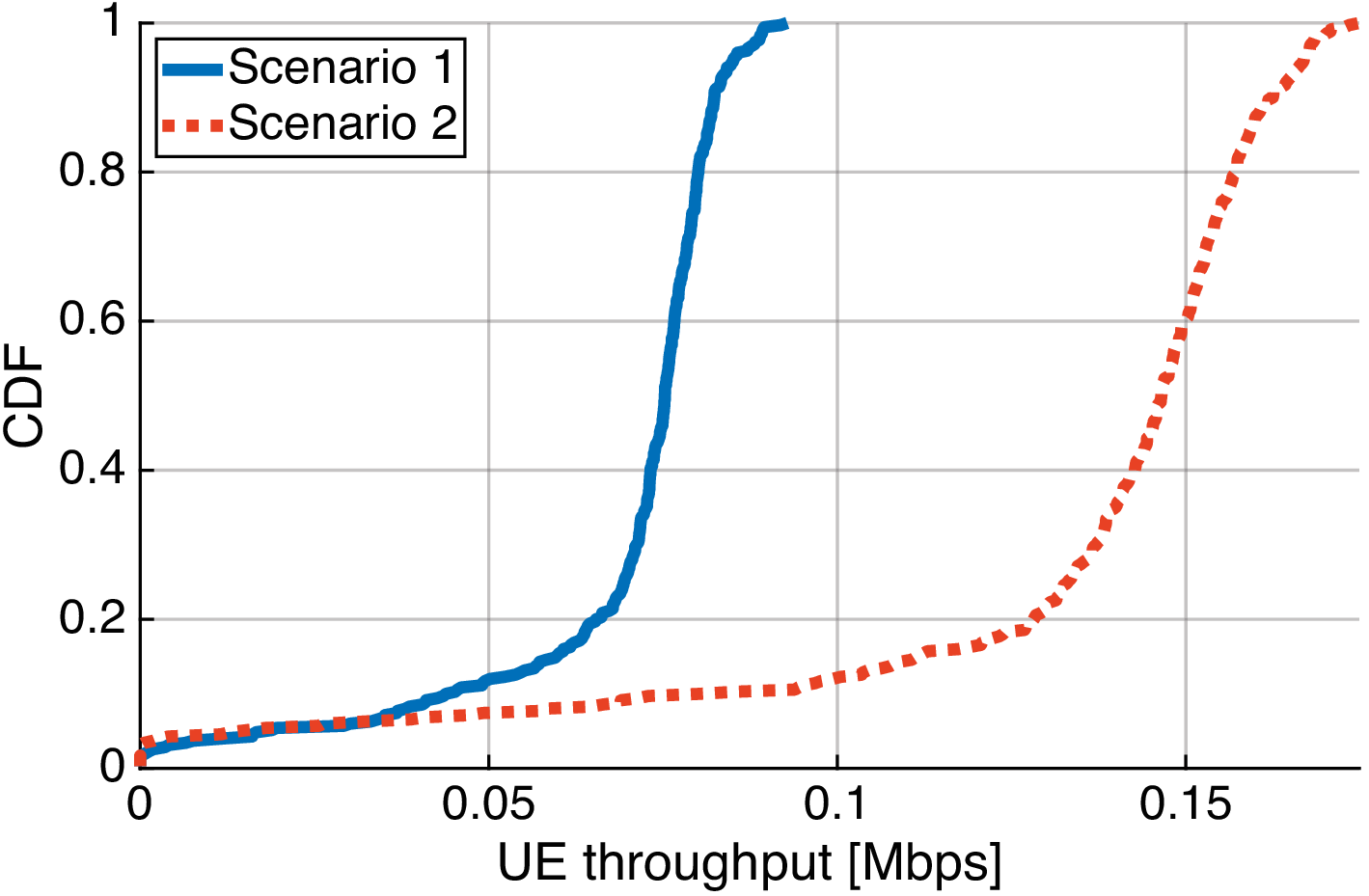}
  }
  \caption{UE throughput in study cases 9 and 10 for Scenarios~1 and~2 under traffic model 4.}
  \label{fig_tm4}
\end{figure}

\subsection{UE Throughput}
Fig.~\ref{fig_tm1to3} shows the cumulative distribution functions (CDFs) of UE throughput for study cases~9 and~10 under traffic models~1 to~3. Each traffic model exhibits a distinct throughput profile that reflects its underlying traffic characteristics.

Traffic model~1 exhibits a consistent throughput scaling trend with respect to packet size. As illustrated in the first column of Fig.~\ref{fig_tm1to3}, the 5th percentile throughput reaches 0.022044~Mbps in Scenario~1 and increases to 0.044089~Mbps in Scenario~2 for study case~9, while corresponding values of 0.023467~Mbps and 0.044089~Mbps are observed for study case~10. This scaling trend extends to the majority of UEs and indicates that comparable throughput levels are achieved under both FRF configurations, with larger packet sizes consistently yielding higher throughput.



Traffic model~2 yields lower throughput than traffic model~1. For example, under study case~9, the 5th percentile throughput is limited to 0.00448~Mbps for Scenario~1 and 0.006471~Mbps for Scenario~2, as illustrated in the second column of Fig.~\ref{fig_tm1to3}, which is substantially lower than the corresponding values observed in traffic model~1. This reduction is attributable to the combination of infrequent packet arrivals and a smaller effective packet size. The smaller effective packet size, together with the lower packet arrival frequency relative to traffic model~1, limit payload transmission efficiency and consequently constrains achievable throughput. A similar tendency is observed under study case~10, where the 5th percentile throughput remains at 0.00448~Mbps in Scenario~1 and increases to 0.007467~Mbps in Scenario~2.



Traffic model~3 shows a distinct throughput profile that is characterized by convergence among higher percentiles. As shown in the third column of Fig.~\ref{fig_tm1to3}, the 5th percentile throughput under study case~9 reaches 0.013867~Mbps in Scenario~1 and 0.020978~Mbps in Scenario~2, whereas the 50th and 95th percentile values converge rapidly to 0.040~Mbps in Scenario~1 and approximately 0.0796~Mbps in Scenario~2. A similar pattern is observed under study case~10, where the 50th and 95th percentile throughputs remain closely aligned at around 0.040~Mbps in Scenario~1 and around 0.0793~Mbps in Scenario~2. The steep rise of the CDFs indicates limited throughput variation among the upper percentiles, which reflects the characteristics of Markov-modulated on/off traffic in which active voice users share resources in a relatively uniform manner.



\begin{table*}[t]
\centering
\renewcommand{\arraystretch}{1.5}    

\newcolumntype{C}[1]{>{\centering\arraybackslash}m{#1}}

\caption{{(a) RB allocation ratio with varying numbers of UEs per beam; (b) Delay distribution at 10 UEs per beam. Simulation is based on the traffic models configured for Scenario 1, as detailed in Table~\ref{table_traffic}.}}

\textbf{(a) Resource block (RB) allocation ratio}\\[0.35\baselineskip]
\begin{adjustbox}{max width=\textwidth}
\begin{tabular}{C{0.15\textwidth} C{0.35\textwidth} C{0.1\textwidth} C{0.1\textwidth} C{0.1\textwidth}C{0.1\textwidth} C{0.1\textwidth}}
\specialrule{1.1pt}{0pt}{0pt}
\rowcolor{myblue!30}

\textbf{Traffic model} & \textbf{Traffic type} & \textbf{Study case} & \multicolumn{4}{c}{\textbf{Number of UEs per beam}} \\[-1pt]
\cline{4-7}
\rowcolor{myblue!30}
& &  & \textbf{10} & \textbf{20} & \textbf{30} & \textbf{40} \\[-1pt]

\specialrule{1.1pt}{0pt}{0pt} 

\multirow{2}{*}{\textbf{1}} & \multirow{2}{*}{\textbf{Poisson process (fixed packet size)}} 
  & 9  &  2.4\% & 5.98\% & 7.72\%  & 12.32\% \\
\cdashline{3-7}
& & 10 &  2.73\% & 9.09\% & 13.85\% & 17.21\% \\
\hline

\multirow{2}{*}{\textbf{2}} & \multirow{2}{*}{\textbf{Poisson process (variable packet size)}} 
  & 9  & 0.54\%  & 1.13\%  & 1.9\% & 2.66\% \\
\cdashline{3-7}
& & 10 & 0.65\%  & 2.14\% & 3.83\% & 4.25\%  \\
\hline

\multirow{2}{*}{\textbf{3}} & \multirow{2}{*}{\textbf{Markov-modulated on/off process}} 
  & 9  & 2.84\%  & 7.38\% & 10.88\% & 15.18\% \\
\cdashline{3-7}
& & 10 & 3.32\%  & 8.07\%  & 16.88\% & 19.91\% \\
\hline

\multirow{2}{*}{\textbf{4}} & \multirow{2}{*}{\textbf{Composite of traffic models {1-3}}} 
  & 9  & 5.87\%  & 12.53\% & 22.71\% & 28.14\% \\
\cdashline{3-7}
& & 10 & 7.03\%  & 18.94\% & 31.34\% & 43.07\% \\
\bottomrule
\end{tabular}
\end{adjustbox}

\vspace{1.0\baselineskip} 

\textbf{(b) Delay distribution}\\[0.35\baselineskip]
\begin{adjustbox}{max width=\textwidth}
\begin{tabular}{
C{0.15\textwidth}   
C{0.35\textwidth}   
C{0.10\textwidth}   
C{0.13\textwidth}   
C{0.13\textwidth}   
C{0.13\textwidth}   
}
\specialrule{1.1pt}{0pt}{0pt}

\rowcolor{myblue!30}
\textbf{Traffic model} &
\textbf{Traffic type} &
\textbf{Study case} &
\multicolumn{3}{c}{\textbf{Delay [ms]}} \\[-1pt]
\cline{4-6}

\rowcolor{myblue!30}
& & &
\textbf{5\%} & \textbf{50\%} & \textbf{90\%} \\
\specialrule{1.1pt}{0pt}{0pt}

\multirow{2}{*}{\textbf{1}} &
\multirow{2}{*}{\textbf{Poisson process (fixed packet size)}} &
9  & 12 & 12 & 119 \\
\cdashline{3-6}
& & 10 & 12 & 12 & 108 \\
\hline

\multirow{2}{*}{\textbf{2}} &
\multirow{2}{*}{\textbf{Poisson process (variable packet size)}} &
9  & 12 & 12 & 236 \\
\cdashline{3-6}
& & 10 & 12 & 12 & 174 \\
\hline

\multirow{2}{*}{\textbf{3}} &
\multirow{2}{*}{\textbf{Markov-modulated on/off process}} &
9  & 12 & 12 & 32 \\
\cdashline{3-6}
& & 10 & 12 & 12 & 32 \\
\hline

\multirow{6}{*}{\textbf{4}} &
\multirow{6}{=}{%
\centering
\textbf{Composite of traffic models 1--3}\\
\footnotesize
(For each study case, the three rows denote the impacts of traffic models 1--3 in order.)
} &
\multirow{3}{*}{9}  & 12 & 13 & 102 \\
\cdashline{4-6}
& &                & 12 & 13 & 56  \\
\cdashline{4-6}
& &                & 12 & 13 & 43  \\
\cdashline{3-6}
& & \multirow{3}{*}{10} & 12 & 13 & 89  \\
\cdashline{4-6}
& &                & 12 & 13 & 68  \\
\cdashline{4-6}
& &                & 12 & 13 & 40  \\
\bottomrule

\end{tabular}
\end{adjustbox}


\label{table_ab}
\end{table*}

Fig.~\ref{fig_tm4} illustrates the CDFs of UE throughput for traffic model~4, which represents a composite traffic scenario in which multiple services are supported simultaneously under service-level priority awareness. In comparison with the individual traffic models, the composite traffic configuration exhibits an overall upward shift in the throughput distribution, which indicates that service coexistence enhances system throughput through increased scheduling diversity and more effective resource allocation. This improvement is primarily observed beyond the lower tail of the distribution. Under study case~9, the 5th percentile throughput remains comparable to that of the individual traffic models, reaching 0.019985~Mbps in Scenario~1 and 0.012947~Mbps in Scenario~2, whereas higher percentiles experience noticeable gains, with the 95th percentile throughput increasing to 0.085760~Mbps and 0.167670~Mbps, respectively. Study case~10 shows a similar trend, in which the 5th percentile throughput is preserved while the upper percentiles continue to improve under the composite traffic model. These results indicate that service coexistence enables throughput gains for the majority of UEs while preserving baseline performance, which is essential for practical multi-service LEO satellite communication systems.


A comparison of study case~9 and~10 across all traffic models in Figs.~\ref{fig_tm1to3} and \ref{fig_tm4} reveals a fundamental tradeoff introduced by frequency reuse. Although study case~10 adopts a higher FRF, which reduces IBI, this benefit is offset by the reduction in per-beam bandwidth. As a result, throughput gains remain constrained, particularly at the lower percentiles. Consequently, improvements in the 5th percentile throughput are limited across all traffic models, which indicates that interference mitigation alone does not yield proportional performance gains for the worst-served UEs.

These throughput distributions offer more than a description of how traffic patterns, packet size, and frequency reuse interact. They provide a quantitative basis for assessing system-level performance guarantees. In particular, the 5th percentile throughput represents the minimum data rate that approximately 95\% of UEs can expect to achieve and therefore serves as a practical measure of baseline service quality that can be used to establish performance floors.


\subsection{RB Allocation Ratio}
Table~\ref{table_ab}a reports the RB allocation ratio as a function of the number of UEs per beam for the four non-full-buffer traffic models under Scenario~1 in study cases~9 and~10.

The evaluation begins with the maximum bandwidth configuration in terms of the number of RBs. According to Table 5.3.2-1 in TS~{38.101-1} \cite{TS38101}, a subcarrier spacing of 15~kHz combined with a channel bandwidth of 30~MHz yields a maximum of 160 RBs. When FRF~$= 3$ is applied, the available transmission bandwidth is evenly partitioned among three reuse groups, leaving one-third of the total RBs available per beam. The RB allocation ratio is then calculated by aggregating the RBs allocated to the central 19 beams over the observation interval, after excluding a 500~ms warm-up period.

As shown in Table~\ref{table_ab}a, the RB allocation ratio rises as the number of UEs per beam grows. This trend reflects the combined effects of higher aggregate resource demand and increased inter-user interference, which together require a larger share of RBs to sustain throughput. With respect to the FRF configuration, study case~10, which employs FRF~$=3$, consistently presents a higher RB allocation ratio than study case~9 with FRF~$=1$. This outcome arises because FRF~$=3$ apportions one-third of the total spectrum to each beam, which reduces the per-beam bandwidth and necessitates the allocation of a larger fraction of the available RBs to accommodate traffic loads, despite the associated mitigation of IBI.

From an operational perspective, the RB allocation ratio directly reflects the level of radio resource utilization at the beam level and thus indicates the system's proximity to overload conditions. As such, it offers a practical metric for network dimensioning and beam-level UE density planning under realistic traffic conditions. Although RB consumption does not scale strictly linearly with the number of UEs per beam reflecting the compounded effects of increased inter-user interference and reduced scheduling efficiency under higher UE densities, the simulation results indicate that the LEO satellite communication systems can accommodate more than 40 UEs per beam while maintaining RB utilization within feasible operational limits across the considered traffic models.

\subsection{Delay}
Table~\ref{table_ab}{b} reports the packet delay distributions for the considered traffic models under Scenario~1 and study cases~9 and~10. These delay characteristics can be evaluated only under non-full-buffer traffic, as the conventional full-buffer assumption maintains persistently backlogged queues and therefore fails to capture packet arrival processes and queue buildup dynamics.

Across all traffic models, the 50th percentile latency remains below 13~ms, which is close to the RTT of the LEO satellite link. For an altitude of 600~km and an elevation angle of 90$^\circ$, the RTT is approximately 12~ms. The RTT considered in this article includes round-trip propagation delays over the service and feeder links, which range between 8 and 9~ms depending on the UE location, as well as a processing and scheduling delay of 3~ms. This observation suggests that at least half of the packets experience minimal queueing delay, with transmissions that are completed without requiring HARQ retransmissions beyond the initial transmission attempt.

By contrast, the 90th percentile delay increases sharply across all traffic models, which indicates that the worst-served 10\% of UEs experience significantly higher latency. Delay behavior is further shaped by the interaction between the service-level priority and the starvation-avoidance mechanism under composite traffic conditions, as observed in traffic model~4. High-priority traffic, which corresponds to traffic model~3, maintains relatively low latency but exhibits an increase relative to its single-traffic baseline, as larger packets from coexisting services introduce head-of-line blocking effects. In contrast, lower-priority traffic, such as traffic models~1 and~2, benefit from the dynamic priority elevation applied to aging packets, which substantially mitigates extreme tail delays. These observations indicate that tail latency is strongly influenced by queue buildup, packet size, and traffic burstiness, as well as by adaptive scheduling mechanisms that operate under service coexistence.

Overall, the delay distributions enable a system-level evaluation of latency-related QoS that is attainable only under non-full-buffer traffic models. Because many practical LEO satellite communication services are subject to explicit latency constraints, the presented results provide a quantitative characterization of the delays that users experience under realistic traffic conditions.

\section{Conclusions}\label{sec5}
LEO satellite communications are consolidating their role as the critical pillar of ubiquitous connectivity, yet reliable performance assessment requires system-level simulation that reflects realistic network operation. Prior studies have largely relied on the full-buffer assumption, which fails to capture the heterogeneous and sporadic nature of practical traffic. This article presented a 3GPP-compliant system-level evaluation of LEO satellite communications under non-full-buffer traffic models, using a proprietary simulator developed in accordance with TR~38.811 and TR~38.821.

System-level performance was examined in terms of throughput, RB allocation ratio, and packet delay. The throughput evaluation showed CDFs across traffic models and identified the minimum data rates achievable by most UEs. The RB allocation analysis captured how resource consumption scales with the number of UEs per beam, providing insight into UE supportability under finite RB budgets. The delay analysis showed that median delays remained close to a single RTT, while the tail of the delay distribution was strongly influenced by queue buildup, an effect that could not be captured under the full-buffer assumption that inherently masks delay variability.

Overall, the system-level simulation mapped the expected LEO satellite performance under realistic traffic conditions and provided a quantitative basis for deployment planning and performance optimization. Future work will extend this study to mobility-induced beam transitions, UL behavior, and multi-satellite constellations that are currently under discussion in 3GPP standardization, toward a more comprehensive system-level characterization of LEO satellite communications.

\bibliographystyle{IEEEtran}
\bibliography{reference}
\end{document}